\documentstyle[aps,pre]{revtex}

\begin{document}

\title{Influence of hydrodynamic interactions on the ballistic deposition
of colloidal particles on solid surfaces.}

\author{I. Pagonabarraga$^{(1)}$\footnote{Current address:
FOM-Institute for Atomic and Molecular Physics, Kruislaan 407, 1098 SJ
Amsterdam,
The Netherlands.}, P. Wojtaszczyk$^{(2)}$, J.M.
Rub\'{\i}$^{(1)}$, B. Senger$^{(3)}$, \\
J.-C. Voegel$^{(3)}$, P. Schaaf$^{(2,4)}$}

\address{(1) Departament de F\'{\i}sica Fonamental, \\
Universitat de Barcelona
	Av. Diagonal 647, 08028 Barcelona, Spain.\\
(2) Institut Charles Sadron,
	6, rue Boussingault, \\
67083 Strasbourg Cedex, France.\\
     (3) Institut National de la Sant\'e et de la Recherche M\'edicale,\\
Unit\'e 424, F\'ed\'eration de Recherches "Odontologie", \\
Universit\'e Louis Pasteur,
	11, rue Human,\\
 67085 Strasbourg Cedex, France.\\
(4) Ecole Europ\'eenne de Chimie, Polym\`eres et Mat\'eriaux de
Strasbourg,\\
	1, rue Blaise Pascal,
	Bo\^{\i}te Postale 296F,\\
 67008 Strasbourg Cedex, France.}

\date{\today}
\maketitle

\begin{abstract}

 The ballistic deposition of particles by taking hydrodynamic
interactions (HI) into account has been studied by means of computer
simulations. The radial distribution function of the assembly of
particles deposited on a plane has been determined as a function of the
coverage and compared to experimental data. It appears that the
introduction of HI in the model when compared to the ballistic model
(BM) predictions leads to a better agreement between experiment and
simulation in particular for the radial distribution function. HI also
modify the value of the first non-vanishing term ($B_3$) in the
expansion of the available surface function ,$\Phi$, in the coverage.
One can estimate the ratio $B_3^{BHM}/B_3^{BM}\approx 0.5$, where
$B_3^{BHM}$ (resp. $B_3^{BM}$) corresponds to simulations in which HI
have been (resp. have not been) taken into account. The introduction of
HI, however, leads to small changes in $\Phi$. Finally, we conclude
that, as far as average global quantities are concerned, the BM without
HI constitutes a good approximation. It is only for the detailed
analysis of the structure of the layer of deposited particles that HI
play a significant quantitative role.

 \end{abstract}


\section{Introduction}

\setcounter{equation}{0}

 Irreversible deposition processes of colloidal particles or
macromolecules on solid surfaces have received considerable attention
during the last years.  Both the adsorption (or deposition) kinetics
and the structure of the assembly of deposited particles have been
analyzed from an experimental and theoretical point of view.  By
irreversible we mean processes in which, once the particle has
interacted with the surface, it can neither desorb from the surface nor
diffuse on it.  Moreover, we will focus on situations in which, due to
the interactions between the spherical particles and the plane, only
one adsorbed layer is formed.  Despite their apparent simplicity, the
deposition processes are determined by the interplay of various
processes:  the Brownian motion of the adhering particle, the
gravitational force, the hydrodynamic interactions (HI) and other kind
of interactions between adhering particles and the adsorbed ones.  Due
to the non additivity of the hydrodynamic interactions, most of the
models which have been developed up to now to describe irreversible
adhesion processes have neglected these latter, and have focused
primarily on the geometric aspects, related to the exclusion surface
effects.  Among all the models which have been developed, two have
captured most of the attention:  (i) On the one hand, the Random
Sequential Adsorption (RSA) model which has been shown to reproduce
some of the properties of the irreversible adsorption of proteins and
small colloidal particles\cite{feder1980}-\cite{wojtaszczyk1995} on
solid surfaces.  By {\it small} it is meant that the motion of the
particles in solution is controlled by Brownian motion, and that
therefore the influence of the deterministic forces on their motion can
be neglected.  In the RSA model particles are placed randomly and
sequentially on the surface.  If an incoming particle overlaps an
already adsorbed one, it is rejected and a new position is chosen
randomly over the surface.  (ii) The Ballistic Model (BM), on the other
hand, has been introduced to account for the deposition of large
particles on solid surfaces\cite{wojtaszczyk1995}-\cite{thompson1992}.
In this case, the deposition process is dominated by gravity and the
diffusion of the particles in the bulk is neglected.  In the BM, again
the position of each incoming particle is randomly selected, but if it
touches an already deposited one, it rolls over this particle according
to the deterministic laws of mechanics.  It undergoes its motion until
it reaches the adhesion plane or is trapped by at least three adsorbed
particles.

     Both models have been compared to experimental observations and the
results can be summarized as follows:\\

\noindent (1) The RSA model has been tested in three different ways:

\begin{enumerate}

\item[a)] Experimental adsorption kinetics of
particles\cite{adamczyk1992} and proteins\cite{ramsden1993} have been
compared to their theoretical expectations. These systems seem to
follow an RSA-like adsorption kinetics at low and intermediate
coverages.  It has also been reported\cite{ramsden1993} that the
jamming limit is reached according to the power law:

\begin{equation}
\theta(\infty)-\theta(t)\sim t^{-1/2}		\label{assrsa}
\end{equation}

     \noindent where $\theta(t)$ represents the coverage after a time t
 of adsorption and $\theta(\infty)$ is the coverage obtained at
saturation.  However, due to the great experimental difficulties to
precisely determine the evolution of the adsorbed amount in the
asymptotic regime, one must be cautious with these results and more
experimental investigations should be performed to validate them.

\item[b)] The statistical properties of surfaces covered by latex
particles have been investigated.  For small particles
depositing under a process widely governed by the diffusion, the radial
distribution function $g(r)$ experimentally found is in good agreement with the
$g(r)$ determined by computer simulations according to the RSA
rules\cite{feder1980}.

\item[c)] Finally, the density fluctuations of adsorbed particles have
also been determined.  Special attention was paid to the reduced
variance, $\sigma^2/<n>$, where $\sigma^2$ corresponds to the variance
of the number of adsorbed particles in sub-systems of a given area out
of the adsorption plane, and $<n>$ represents the mean number of
adsorbed particles on these sub-systems.  It can be shown that
$\sigma^2/<n>$ is directly related to the radial distribution function
$g(r)$ through the relation:

\begin{equation}
\frac{\sigma^2}{<n>}=1+\rho\int_0^{\infty}2\pi r \left[ g(r)-1 \right] dr
					\label{siggr}
\end{equation}

     \noindent where $\rho=<n>/S$, $S$ being the area of the adsorption
plane\cite{landau1959}.  Surprisingly, it has been found that, for the
systems which have been investigated, $\sigma^2/<n>$ does not follow
the RSA behavior as a function of $\rho$ (or $\theta$) but behaves more
closely to the BM predictions\cite{schaaf1995,mann1995}.  This result
can be explained as follows:  if one expands $\sigma^2/<n>$ as a
function of the coverage :

\begin{equation}
\frac{\sigma^2}{<n>}=1-B_n\theta^n+O(\theta^{n+1}),
					\label{sigexp}
\end{equation}

\noindent the order $n$ of the first non vanishing term $B_n\theta^n$
represents the smallest number of particles which are required on the
surface in order to hinder the adhesion of a new particle.  For
example, in the RSA case $n=1$, and in the BM case $n=3$.  Moreover,
$B_n$ is related to the mean exclusion area of $n$ deposited
particles.  In the case experimentally investigated, which led to an
RSA-like radial distribution function, the Brownian motion of the
particles in the bulk largely dominates over the gravitational effects
even if this latter is not totally absent.  So, a diffusing particle
from the bulk that will touch an already adsorbed particle will diffuse
around it at distances which are large compared to the radius of the
particles.  However, due to the slight gravitational effect, it will
finally reach the adsorbing surface.  Thus, in this case, $n$ must
exceed 1: one deposited particle cannot hinder another particle to
adhere on the surface.  On the other hand, the presence of HI favors
the diffusion of the particle parallel to the plane, implying that the
incoming particle can explore much larger distances along the surface
than perpendicularly to it before adhering\cite{bafaluy1993}.  As a
consequence, in first approximation, the incoming particle can be
assumed to adsorb randomly on the surface.  This implies that the
radial distribution function becomes very close to the $g(r)$ predicted
by the RSA model.  These slight differences between the RSA-like $g(r)$
and the experimental radial distribution function are responsible for
the totally different behavior of the experimental $\sigma^2/<n>$ as
compared to the RSA one.  This observation shows how subtle the
deposition process can be and the importance of the detailed
description of the transport process from the bulk to the interface, in
order to account properly for the statistical properties of such
assemblies of spheres.

\end{enumerate}

\vspace{0.5cm}

\noindent (2) For the deposition of large particles on surfaces in the
absence of shear at the interface only the statistical properties of
the assembly of spheres have been experimentally determined.  A good
agreement has been found in this case between the evolution of
$\sigma^2/<n>$ as predicted by the BM and the experimental data.  This
result is expected if one refers to our discussion in the RSA
case\cite{schaaf1995}.  The radial distribution function $g(r)$ has
also been determined as a function of the
coverage\cite{wojtaszczyk1993}.  Despite the fairly good agreement
between experimental data and simulations, some discrepancies remain,
especially at distances slightly larger than one particle diameter.
Such discrepancies have first been attributed to some polydispersity in
the particle sample.  However, even if the introduction of
polydispersity leads to a better agreement between experiment and
theory, it cannot account for the whole
differences\cite{wojtaszczyk1993,wojtaszczykth}.  It has thus been
proposed that these differences are due to the hydrodynamic interactions
between the incoming particle, the adsorbed particles and the adsorbing
surface.  Indeed, whatever the radii of the particles, the HI are
always present during the deposition process and their relative
importance compared to the gravitational forces does not decrease when
the radii of the particles increase.  Moreover, the HI are of long
range and can thus have significant effects on the distribution of the
particles on the deposition plane.

     It is the purpose of this article to address the problem of the
influence of the HI during the irreversible
deposition of large colloidal particles on the statistical properties
of the assembly of the deposited particles.  The study will be
performed by means of computer simulations, extending to the 3-D case
the study of the effect of HI on the deposition on a one-dimensional
substrate\cite{ignacio1994}, and the results will be compared to
experimental data.  We will first present the theoretical background
which is necessary to be able to compute the friction tensor which
determines the motion of the depositing particles.  We will, in
particular, describe the approximation which will be made in our
approach.  In the next section we will present the simulation model.
We will then show the results related to the radial distribution
function of the particles on the surface.  The $g(r)$ will be compared
to the radial distribution function corresponding to the BM model in
which HI are neglected, and to experimental results.  The available
surface function, $\Phi$, which represents the adhesion probability of
a particle on the surface, will also be discussed and compared to its
counterpart in the BM case.  Particular attention will be paid to the
third virial coefficient, $B_3$, in the development of $\Phi$ in a
power series of $\theta$. The coefficient $B_3$ is related to the area
associated with the triangles of preadsorbed spheres which trap
incoming particles, preventing them from being adsorbed.  A new
simulation method aimed to determine the value of $B_3$ will be
presented.  Finally, we will summarize our results and draw some new
lines along which new investigations should be performed. In the
appendix, explicit expressions for the components of the friction
tensor for two spheres suspended in an unbounded fluid are given for
the sake of completeness.

\section{Hydrodynamic interactions in the adsorption process}
				\label{hydro}

	The slow motion of a particle suspended in a fluid at rest is
governed by the frictional force and torque that the fluid
exerts on it. These are linear functions of the center of mass
velocity, $\vec{u}$, and angular velocity, $\vec{\omega}$ of the particle,
the proportionality coefficients being the so-called friction tensors.
One can therefore write down:

\begin{eqnarray}
\vec{F}&=&-\vec{\vec{\xi}}_{tt}\cdot\vec{u} -
\vec{\vec{\xi}}_{tr}\cdot\vec{\omega}			\nonumber\\
\vec{T}&=&-\vec{\vec{\xi}}_{rt}\cdot\vec{u}-
\vec{\vec{\xi}}_{rr}\cdot\vec{\omega}			\label{1.1}
\end{eqnarray}

\noindent where  $\vec{F}$ and $\vec{T}$ account for the total force
and total torque acting on the particle.  $\vec{\vec{\xi}}_{tt}$ and
$\vec{\vec{\xi}}_{rr}$ correspond to the translational and the
rotational friction tensors respectively, while $\vec{\vec{\xi}}_{rt}$
and $\vec{\vec{\xi}}_{tr}$ are the coupling friction tensors.  Due to
the reciprocal Onsager relations, they satisfy
$\vec{\vec{\xi}}_{rt}=\vec{\vec{\xi}}_{tr}^{ \dag}$, where the superscript
$\dag$ indicates that the transpose of the matrix should be taken.

	 We are interested in the motion of a free spherical particle in the
presence of the gravitational field.  Due to the symmetry and homogeneity
of the particles, there will be no net external torque acting on them.
Therefore $\vec{T}=0$, and from eqs.(\ref{1.1}) one can deduce the
appropriate expression for the center of mass velocity, which will
completely determine the motion of the particles.  One then obtains

\begin{equation}
\vec{F}=-\left(\vec{\vec{\xi}}_{tt} - \vec{\vec{\xi}}_{tr} \cdot
\vec{\vec{\mu}}_{rr} \cdot \vec{\vec{\xi}}_{tr}^{ \dag}\right) \cdot
\vec{u}= -\vec{\vec{\xi}}_{eff}\cdot \vec{u}
								\label{1.2}
\end{equation}

\noindent where $\vec{\vec{\mu}}_{rr}$ is the rotational mobility matrix,
defined as the inverse of the corresponding friction tensor,
$\vec{\vec{\mu}}_{rr} \cdot \vec{\vec{\xi}}_{rr} = \vec{\vec{1}}$.
The coefficient between
brackets in eq.(\ref{1.2}) can be redefined as an effective friction
tensor, $\vec{\vec{\xi}}_{eff}$.  The relaxation times for the velocity
of colloidal particles are very small, which means that the inertial
terms in their motion can be neglected. This implies that the
hydrodynamic force $\vec{F}$ that the fluid exerts on the particle is
exactly opposite to the external force, which in our case is the
gravity force acting on the suspended particles.  Moreover, since we
are interested in the deposition of heavy colloidal particles, the
effect of Brownian diffusion on  their motion can be neglected.
Therefore, eq.(\ref{1.2}) constitutes the equation of motion of the
suspended particles, putting $\vec{F}=-\frac{4}{3}\pi a^3 \Delta\rho g
\hat{z}$ the gravity force, where $a$ is the radius of the sphere, $g$
the gravity acceleration, and $\Delta\rho$ the density difference
between the colloidal particle and the solvent, assumed to be a positive
number.

	Eq.(\ref{1.2}) will describe the motion of a suspended heavy
particle both in the case in which it is alone in an unbounded fluid
and in the presence of other objects.  The HI
 between the different particles, which are mediated by the host fluid,
appear through the specific expressions for the friction coefficients.
These depend both on the geometry of the particles and on their
relative distribution in the fluid.  For example, for an isolated solid
sphere, the friction coefficients are constants, and using the standard
stick boundary condition for the velocity field on its surface one has,
$\vec{\vec{\xi}}_{tt}=6\pi\eta a\vec{\vec{1}}\equiv \vec{\vec{\xi}}_0$ and
$\vec{\vec{\xi}}_{rr}=8\pi\eta a^3\vec{\vec{1}}$ with $\eta$ being the
viscosity of the solvent.  In this case, there exists no coupling
between translational and rotational motion, $\vec{\vec{\xi}}_{tr}=0$.

	 We will be interested in geometries where one sphere
approaches a planar surface covered by preadsorbed spheres.  In this
case, there exist no general analytic expressions for the friction
coefficients since exact solutions for many-body hydrodynamic problems
are not available.  The difficulty in the derivation of an exact
solution lies partially in the non-additive character of the HI, which
arises as a consequence of the long-range decay of the Oseen
propagator\cite{happel}.  However, in order to get analytic
expressions, additivity in the HI will be assumed.  There are two
natural ways to introduce such an approximation:  one can assume
additivity of the friction tensors, which is equivalent to assume that
the hydrodynamic force acting on one particle due to the presence of
the others is equal to the sum of the forces due to each one of them as
if the others were not present.  One can also assume additivity of the
mobilities, according to which the velocity of a given particle is
equal to the sum of the velocities induced on it by the other particles
independently.  Bossis and Brady\cite{bossis1984} have shown that the
additivity of the friction coefficients takes properly into account the
lubrication forces, which act when objects are close together.  This
latter property is not fulfilled under the mobility additivity
assumption.  Within the friction additivity assumption, the effective
friction tensor for a suspended particle at a height $z_v$ from a
planar surface in the presence of $N$ previously adsorbed spheres will
be written as

\begin{equation}
\vec{\vec{\xi}}(\vec{r})_{eff}=\vec{\vec{\xi}}_{sp}(z_v)+\sum_{i=1}^N \left(
\vec{\vec{\xi}}_{ss}(\vec{r}_i)-\vec{\vec{\xi}}_0\right)
							\label{1.3}
\end{equation}

\noindent where $\vec{r}$ represents the position vector of the
incoming particle with respect to a given reference frame (we will take
the origin of the reference frame at the center of the closest
preadsorbed particle), $\vec{r}_i$ is the vector joining the center of
the incoming particle and of the $i$th sphere on the plane,
$\vec{\vec{\xi}}_{sp}$ is the friction tensor of a spherical particle
alone in the presence of a plane, $\vec{\vec{\xi}}_{ss}$ the effective friction
tensor of two spheres in an unbounded fluid at relative position
$\vec{r}_i$, and $\vec{\vec{\xi}}_0$ is the
Stokes' tensor, introduced in the preceding paragraph.
  The previous expression leads to the correct behavior
for the friction tensor when the sphere is far from the surface, where
one recovers Stokes law.  On the other hand, when the adsorbing
particle comes into the vicinity of an already adsorbed one,
lubrication forces become dominant.  Then, only the small region of the
fluid between the particles is responsible for the forces which can,
thus, be considered as additive.  Therefore, eq.(\ref{1.3}) also
provides the right behavior at short distances.  In addition, the
advantage of introducing approximation (\ref{1.3}) is that the
expressions for the friction coefficients appearing in it are known.
Indeed, the friction tensor for a sphere at a height $z$ in the
presence of a plane, $\vec{\vec{\xi}}_{sp}$, has been shown to be
diagonal and to have one component parallel, $\xi_{||}$, to the plane
and one perpendicular to it, $\xi_{\bot}$
\cite{brenner1961,brenner1967}:

\begin{eqnarray}
\xi_{\bot}&=& \frac{4}{3}\sinh \alpha \sum_{k=1}^\infty \frac{k (k+1)}{(2
k-1)
(2 k+3)}  \left( \frac{2 \sinh ((2 k+1)\alpha)+(2 k+1) \sinh 2\alpha}{4
\sinh^
2 ((k+1/2)\alpha)-(2 k+1)^2 \sinh^ 2 \alpha} -1\right)  \label{1.4} \\
\xi_{||}&=&\left\{\begin{array}{lll}
\left(1-\frac{9 a}{16 z}+\frac{1 a^3}{8 z^3}- \frac{45 a^4}{256 z^4}-
\frac{a^5}{16 z^5}+...\right)^{-1}&  &   \;\;\;\;\;\;\frac{z-a}{a}>0.1 \\
 & & \\
-\frac{8}{5}\ln \left(\frac{z}{a}-1\right)+.95888+...&  &   \;\;\;\;\;\;
			\frac{z-a}{a}<<0.1
\end{array}
\right.						\label{1.5}
\end{eqnarray}

\noindent with $\alpha=$arccosh $z/a$. It is worthnoting that at small
particle-plane distances, $\xi_{||}$  diverges much more slowly than
$\xi_{\bot}$. Since the diffusion coefficient behaves as the inverse of the
friction tensor, this fact is responsible for the "{\it randomization}" of the
final position of an adhering sphere whose motion is controlled by Brownian
motion, as commented in the introduction\cite{bafaluy1993}. Regarding the
expression for $\xi_{ss}(\vec{r}_i)$, we will use the friction coefficient of
two
spheres in an unbounded fluid as worked out at all
distances by Jeffrey and Onishi \cite{jeffrey1984}. According to these authors,
the effective friction
tensor for two spheres at a distance $\vec{r}$ in an  unbounded fluid is
given by:

\begin{equation}
\vec{\vec{\xi}}_{ss}(\vec{r})=X_{11}^A(\vec{r}) \hat{e}\hat{e}+\left(
Y_{11}^A(\vec{r})-\frac{\left(Y_{11}^B(\vec{r})\right)^2}{3
Y_{11}^C(\vec{r})}\right)
(\vec{\vec{1}}-\hat{e}\hat{e})
							\label{1.6}
\end{equation}

\noindent $\hat{e}=\vec{r}/r$ stands for a unit vector along the line of
centers
of
the two spheres, and the expressions for the functions $X_{11}^A$, $Y_{11}^A$,
 $Y_{11}^B,$ and $Y_{11}^C$ are given in the appendix for the sake of
 completeness.

	Expression (\ref{1.3}), when inserted in (\ref{1.2}), provides
us with the equation of motion of a suspended particle,

\begin{equation}
\frac{d \vec{r}}{d t}= \vec{\vec{\mu}}_{eff}\cdot\vec{F}_g
							\label{1.7}
\end{equation}

\noindent where where the effective mobility $\vec{\vec{\mu}}_{eff}$ is
the inverse of the effective friction tensor given in eq.(\ref{1.3}).

	The non-linear dependence of the friction tensors as functions
of the distance between the particles, together with the fact that the
friction tensors associated with the sphere-sphere configuration have a
spherical symmetry different from the cylindrical one characteristic of
the sphere-plane friction tensors, make it impossible to find an
analytic solution for eq.(\ref{1.7}) even in the simplest case in which
a single sphere is adsorbed on the plane.  In the general situation of
adsorption kinetics, one has to take into account that aside from this
problem, the number of particles on the surface increases with time,
which implies that only a numerical simulation study of the process
can be carried out.

\section{Simulation model}		\label{numerical}

	We have performed numerical studies simulating the trajectories
of the colloidal particles from the bulk to the surface, taking into
account that as soon as a moving particle touches the surface, it is
irreversibly fixed at that position. The model that we will develop
constitutes a natural extension of the Ballistic Model. We will study
the arrival of particles from the bulk to the surface in a sequential
way.  This corresponds to the physical situation in which the bulk
concentration is low so that the interactions between the particles in
the bulk can be neglected.  This situation is indeed encountered in the
experimental systems to which we will refer later on.

	In our simulation algorithm, a position is randomly chosen at a
height of 10 particle diameters above the plane. This ensures that
initially the effects of the interactions of the incoming sphere and
the preadsorbed ones can be neglected.  In all the simulation studies,
we have rescaled the distances so that the diameter of the spheres is
taken as unity, and the time is rescaled by the characteristic
sedimentation time $9 \eta /(2 a \Delta \rho g)$. With this
adimensionalization, when the spheres are far from the plane, their
mobility is equal to 1. This scaling, which is possible due to the
structure of eq.(\ref{1.7}), implies that our results are of general
validity and will not depend on the particular system we consider, as soon
as both
inertial and  diffusion effects are negligible.

We have numerically evaluated the trajectories of the incoming
particles by numerically integrating eq.(\ref{1.7}) with the expression
for the friction tensor as given by eq.(\ref{1.3}), until they reach
either the surface or one or a set of preadsorbed particles. The
integration is performed by using a 4th order Runge-Kutta algorithm
\cite{abramowitz1972} with variable time step. Since the friction
tensor depends on the position of the adsorbing particle relative to
the preadsorbed ones, at each time step it is necessary to calculate
the appropriate value of the tensor.  Moreover, each time a particle
adheres, it has to be taken into account in the evaluation of the
friction tensor of subsequent adhering spheres.  In the numerical
integration it is important to take into account that the behavior of
the friction changes qualitatively along the trajectory of the
particle.  When the sphere is far from any other object, the mobility
is of order unity, and changes slowly.  As the particle comes close to
an adsorbed sphere, the mobility becomes anisotropic.  Then, while the
component associated with the displacement along their line of centers
vanishes linearly with the clearance between the spheres (see
eq.(\ref{1a.4})), the component related to the displacement at constant
separation goes to zero as the inverse of the logarithm of the
clearance (see eqs.(\ref{1a.5})-(\ref{1a.7})). Thus, in this region,
the mobility changes rapidly, and in a different manner depending on
the direction.  In fact, the motion will consist basically of angular
displacements at practically constant distance between the spheres.
Therefore, the variable time step is chosen to ensure that the
displacement is never larger than a tenth of the clearance (see fig.
1).  Moreover, in order to take the anisotropic behavior of the
mobility into account, eq.(\ref{1.7}) is solved in spherical
coordinates centered on the adsorbed sphere closest to the incoming
one.

     The fact that the mobility goes to zero when the spheres come into
     contact, due to the stick boundary conditions, introduces an
additional computational difficulty.  Indeed, the velocity in that
region can become so small that the computer time needed to describe
the trajectory becomes exceedingly large.  We have decided to stop the
trajectory when the clearance between the incoming and an adsorbed
sphere becomes smaller than $10^{-4}$ particle diameters.  For
particles of diameter $2\mu$m, this corresponds to a minimum clearance
of 200 $\AA$.  At this point, we impose that the particle will follow
the steepest descent path towards the surface.  Accordingly, we have
implemented the BM algorithm\cite{thompson1992,choi1993} to
calculate the final position of the sphere on the plane, if it is not
hindered to reach the plane by a group of preadsorbed particles.  This
assumption seems reasonable because at that point, the trajectory is
basically controlled by the geometrical constraint that spheres cannot
overlap, while the external force drives the particle to the surface.
If the particle cannot reach the plane, it is rejected. We thus neglect
multilayer effects.

     When the sphere comes close to the plane, the mobility becomes
anisotropic and exhibits the same qualitative behavior as explained in
the previous paragraph.  Again, the mobility tensor goes to zero when
the clearance vanishes. Therefore, we also change the time step in
order to ensure that the displacement is smaller than a tenth of the
clearance between the incoming sphere and the plane. Moreover, in this
case we have also to stop the numerical integration of the trajectory.
We have considered that when the gap between the sphere and the plane
is smaller than $10^{-2}$ diameters, the particle is deposited on the
surface at that position.  This truncation procedure can be seen as an
effective way to account for the attractive short-range
particle-surface potential, which binds the sphere to the substrate.
The situation would be completely different for trajectories controlled
by Brownian motion.  The sphere would then have a large tendency to
diffuse parallel to the line, leading to a randomization in its final
position on the substrate\cite{bafaluy1993}.

	A second feature which should be taken into account in the
numerical algorithm is the long-range character of HI.  In fact, in
order to obtain the expression of the effective friction tensor,
eq.(\ref{1.3}), a sum over all previously adsorbed spheres should be
carried out.  However, from the computational point of view such a
procedure is time consuming. Therefore, a compromise should be reached
between the number of preadsorbed particles which will be considered to
compute the friction tensor and the computer time needed.  We have then
used the results obtained from the study of the deposition of particles
on a one dimensional substrate in the presence of
HI\cite{ignacio1994}.  In this case, it has been shown that if the
in-plane initial separation to a preadsorbed sphere is of the order of
10 particle diameters, the effect of HI is negligible on the final
location of the incoming sphere. This fact suggests that we can now
restrict the interaction of the incoming spheres to all the preadsorbed
ones lying in a cylinder of radius 10 diameters centered on the
incoming particle (see fig.  2).  Moreover, as the particle approaches
the surface, the value of the friction tensor will be dominated by the
adsorbed particles closest to the adhering one.  We have therefore
further restricted the range of interaction by considering that the
incoming sphere will only be affected by those particles whose
distance, $d$, to the projection of the incoming particle on the plane
is smaller than the height, $h$, at which this incoming sphere is
located.  This restriction procedure, however, is only taken into
account when the height of the incoming particle is larger than 5.  At
lower heights, the radius of the {\it interaction cylinder} is assumed
constant and equal to 5 (see fig.  2).  This further conjecture is
based on the observation that the deviation from the straight
trajectory of the incoming sphere as determined by the gravity field
takes place basically at distances of 2 or 3 over the
plane\cite{ignacio1994}.  Although this reasoning is based on
trajectories obtained with a small number of particles deposited on the
surface, it seems reasonable that it also holds in more complex
geometries.  The advantage of this approximate procedure is that it
saves a significant amount of computer time with respect to the initial
cylindrical restriction.

     We have checked the errors induced by the use of the varying {\it
interaction cylinder} by covering a surface both using a
constant-radius and a varying-radius interaction cylinder, as shown in
figure 3.  At an intermediate coverage, as shown in figure 3 a,
practically all the particles end at the same positions on the plane
using both methods, whereas at higher coverages a small fraction of
the particles are placed at different positions, as seen in figure 3
b.  This is due to the fact that if an incoming sphere arrives close to
an ensemble of preadsorbed spheres, a small initial deviation may lead
to a completely different final position.  Then, due to the infinite
memory of the adsorption process, all the particles arriving afterwards
will be sensitive to this difference, leading eventually to a different
configuration on the surface.  Although {\em a priori}, from the point
of view of average quantities, it is not clear whether this change may
lead or not  to significantly different statistical properties of the
surface, comparison of the data  obtained by this procedure with the
experimental ones shows {\it a posteriori} that this approximation works
 pretty well, since  no significant differences
are observed.

     We have performed numerical simulations of the adsorption process
 on a rectangular surface, of sides 23.34 and 27.34, up to a coverage
of $\theta=0.5$.  The size of the system has been chosen such that the
ratio "size of the system/size of the particles" is the same in the
experiments and the simulations (see also next section).  We have
focused on the study of the radial distribution function, $g(r)$, and
of the available surface function, $\Phi (\theta)$, since both
represent key quantities for the statistical properties of the adsorbed
layer. We have stopped our simulations at a coverage of 0.5, which
corresponds to 407 particles deposited on the surface.  At this
coverage, one can already have an idea of the behavior of the system
near jamming, without having to reach it.  Indeed, the study of the system
in the last 10\% until jamming constitutes computationally the
most expensive part by far.

     We have covered 200 surfaces, and from them, we have constructed
 the radial distribution function.  Simulations of the adsorption
process according to BM rules have also been carried out following
Ref.  \cite{thompson1992} , under the same conditions as those exposed
in the previous paragraph, in order to compare both models. The model
described in this section, which takes HI into account will be referred
to as BHM algorithm hereafter.

\section{Results and discussion}        \label{results}

     All the simulation results were compared with experimental data.
 The latter correspond to the irreversible deposition of melamine
particles of diameter 4.5 $\mu$m and density 1.5 g/cm$^3$ on a silica
surface.  The preparation of the particles, their characterization and
the experimental procedure were reported extensively in reference
\cite{wojtaszczyk1993} and will only be briefly mentioned here.  The
experimental system consists of a cell, a flow system that allows the
injection of the solution in the cell, an inverted microscope (Axiovert
10, Zeiss, Germany), a CCD video camera (type 4710 CCIR monochrome from
Cohu, San Diego, CA) and a computer image analysis system (Visilog,
Noesis, France).  The experiments were performed in a plane parallel
cell whose top and bottom parts are constituted by two silica
microscope slides.  Once the particle solution was introduced in the
cell, it was fixed in its horizontal position to allow the particles to
deposit on the lower microscope slide.  After the deposition of all the
particles present in the cell, a large number of pictures from
different regions of the bottom surface were taken.  Typically, for
coverages of the order of 10-20\%, 300 pictures were needed to obtain
satisfactory results from the subsequent image analysis.  The area that
is covered by each picture equals s = 105 x 123$\mu$m$^2$.  Only the
particles that touch the surface were taken into account, the other
particles, which can form a second layer were discarded by the image
analysis.  The geodesic center of all the particles was then determined
for each picture.  From this set of data, the different statistical
properties of the surfaces such as the radial distribution function
$g(r)$ and the reduced variance of the density fluctuations could be
determined.  The details of the experimental method are given in
reference \cite{wojtaszczyk1993}.  It must be pointed out that great
care had to be taken to determine the $g(r)$ due to the discrete nature
of the positions of the particles determined experimentally because of
the finite size of the pixel elements in the camera.  The reduced
radius $R^*$ of the particles was equal to 3.4.  Let us recall that $R*$
is defined by $R^* = a (4\pi \Delta \rho g/(3 k T))^{1/4}$ , where $a$
corresponds to the radius of the colloidal particles, $\Delta\rho$ to
the density difference between the particle material and the solvent
and $kT$ the thermal energy.  In the absence of HI, this is the only
parameter which is relevant for the description of the deposition
process\cite{senger1992}.  It has been shown in reference
\cite{wojtaszczyk1993} that this system behaves in first approximation
in a ballistic way, even if systematic deviations from the model are
observed, e.g.  in  $g(r)$. However, the experimental data have never
been compared to simulations of deposition processes in which the HI
are taken into account. This constitutes the main objective of the
present work.  In order to assess the characteristic features due to
the HI we have also compared our results with simulations performed in
the framework of the pure BM.

     From the previous studies it comes out that the main parameters
describing the structure of the assembly of deposited particles are the
radial distribution function $g(r)$ (section \ref{gr}), the available
surface function $\Phi(\theta)$ (section \ref{phitheta1}) and the
reduced variance of the density fluctuations of the number of adsorbed
particles $\sigma^2/<n>$.  These two quantities do contain the same
information at low coverage which is quantified by the third virial
coefficient $B_3$ in the expansion of $\Phi$\cite{schaaf1995} and to
which special attention will be paid in section \ref{phitheta2}.

     \subsection{Radial distribution function}	\label{gr}

     The radial distribution function $g(r)$ characterizes the
correlation between the adsorbed particles, and it can be determined
from the positions of the different particles on the surfaces.
However, in order to be able to quantitatively compare the $g(r)$
obtained from the simulation with the experimental ones, it is
necessary to treat the simulation data exactly in the same way as the
experimental ones.  In particular, we have to apply the pixelization
procedure (discretization of the positions of the particles) to the
coordinates of the particles deposited on a surface by means of the BM
or the BHM algorithms.  Taking into account that the diameter of the
melamine particles used in the experiments is 4.5$\mu$m, the dimensions
of the experimental pixel ({\em i.e.}  0.48$\mu$m x 0.41$\mu$m) have
been scaled to 0.48 / 4.5 and 0.41 / 4.5, along the x- and y-axis,
respectively. After performing this rescaling, the unit of distance is
the diameter of the particles. The coordinates of the positions of the
centers of the particles are  then converted into integer numbers of
pixels. The center-to-center distances are calculated using these
integer coordinates, and the histogram for the relative distribution of
particles is evaluated with a width resolution not larger than the
smallest side of a pixel ({\em i.e.}  0.41 / 4.5) as was done with the
experimental data\cite{wojtaszczyk1993}.  This is repeated over the 300
to 500 surfaces available (as indicated in the figure captions).  The
radial distribution function is finally deduced from this cumulated
distance frequency histogram.

     The results are shown on figure 4 (a-f).  For a coverage $\theta$
 of about 0.15, only the first peak can be clearly identified (fig.  4
a,b).  The introduction of HI in the description of the deposition
process when compared to the BM model results mainly  {\it (i)} in the
lowering of this contact peak, which seems then to be in better
agreement with the experimental data, and {\it(ii)} in the broadening
of the peak after its maximum, due to the effective repulsion induced
by HI between the adsorbing and adsorbed spheres.  At intermediate
coverages, $\theta\approx$ 0.35 (fig.  4 c,d), the difference in the
height of the peak is less marked; however, the agreement between the
simulated data and its experimental counterpart is excellent, and they
almost coincide in the
broaden region and becomes better in the region around $r /(2 a) =
1.5$.  This observation may also be done in the case of high coverages
$\theta\approx$ 0.50 (fig.  4 e,f).  However, for this latter coverage,
the simulated data are quite similar, whether or not the HI are taken
into account. This is due to the fact that in this regime, the arrival
of particles at the surface is almost entirely controlled by
geometrical restrictions, since the available area is then only formed
by small targets.  Therefore, the fraction of rolling
particles, which determines the height of the peak, is not so sensitive
to HI repulsion because then the arriving particle is forced to enter
in one of these target areas and the repulsive friction forces are
greatly balanced due to the number of particles surrounding each hole.
 As a general conclusion, HI do only introduce slight changes in the
structure of $g(r)$. They induce, in particular, an effective repulsion
between the particles. The most important change is observed in the
first peak, because, at low coverage, it is related with the rolling
mechanism over one particle, which is indeed very sensitive to
HI\cite{ignacio1994}.

\subsection{Available surface function}	\label{phitheta}

     \subsubsection{Analysis over the entire coverage range}
					\label{phitheta1}

     In order to further analyze to what extent the introduction of the
hydrodynamic forces modifies the structure of the particle
configurations on the adsorbing surface, we present in this section the
comparison of the available surface function $\Phi(\theta)$
corresponding to the BM and the BHM, and denote it by
$\Phi^{BM}(\theta)$ and $\Phi^{BHM}(\theta)$ respectively.  For a given
coverage $\theta$, this quantity is equal to the probability that a
deposition trial will be successful.  For the ballistic deposition, it
is theoretically known that $\Phi^{BM}(\theta)$ behaves as
$1-B_3^{BM}\theta^3+O(\theta^4)$. $B_3$ has first been estimated to
9.61205 by Thompson and Glandt\cite{thompson1992}, and later on
corrected to 9.94978 by Choi {\em et al.}\cite{choi1993}.  It is worth
noting that, in fact, $1 - B_3^{BM} \theta^3$ is an acceptable
approximation only over a narrow coverage range (up to 10 or 15\%).
The absence of the first and second order terms from the series
expansion of $\Phi^{BM}(\theta)$ is due to the fact that the presence
of at least three absorbed particles is required for an incoming
particle to be rejected.  Indeed, a new spherical particle can always
roll over one or two fixed spheres and reach eventually the surface.
This is also true when HI are involved in the deposition process.
Hence, the expansion of $\Phi$ as a power series of $\theta$
corresponding to the BHM model must be of the form
$\Phi^{BHM}(\theta)=1 - B_3^{BHM} \theta^3+O(\theta^4)$ with
$B_3^{BHM}$ {\em a priori} not equal to $B_3^{BM}$.

     The values of $\Phi(\theta)$ derived from the simulations are
shown on figure 5a, where we compare the data obtained in the framework
of the BM (400 surfaces) and of the BHM (500 surfaces up to $\theta$ =
0.3, and 300 surfaces from $\theta$= 0.3 up to $\theta$ = 0.5).
Without any additional computation it is obvious that both data sets
are not identical, even though the sample sizes lead to a non
negligible noise level.  If we fit a polynomial of the fifth degree
(without the $\theta$- and $\theta^2$- terms) to the simulation data
over the whole range (coverage from 0 to 0.5), we obtain
$B_3^{BM}\approx 9.427$ and $B_3^{BHM}\approx 4.695$.  Even though the
value $B_3^{BM}$ is not exactly identical to its theoretical prediction
(9.94978), it clearly appears that the third-order coefficient is
strongly influenced by HI ( $B_3^{BM}/B_3^{BHM}\approx 2$).  This
reduction indicates that the deposition probability falls off less
rapidly when HI are introduced in the model.  These values must however
be taken with great care due to the poor statistics and to the possible
mutual influence of the fitting parameters.  Moreover, it must be
realized that using the values of the fits for the available surface
function leads to the values $\Phi^{BM}(\theta = 0.1)$ = 0.9906 and
$\Phi^{BHM}(\theta = 0.1) = 0.9953$, which are almost equal.  This
confirms the observation following from the comparison of the $g(r)$
(see preceding section) that HI do not deeply alter the deposition
process on global average magnitudes in the present conditions.
Nonetheless, an additional more precise simulation has been performed,
in which 5000 surfaces were covered up to a coverage of 0.1 (fig. 5b).
In this regime $\Phi^{BHM}$ ( as $\Phi^{BM}$) should be accurately
expressed by $1-B_3 \theta^3$. The value of $B_3^{BHM}$ deduced from
these data is 4.849, which is not far from the former estimate derived
from the full data set shown in fig. 5a.

     Figure 6 renders clear that the difference between $\Phi^{BHM}$
and $\Phi^{BM}$ is almost everywhere positive.  This systematic
character of the sign of $\Phi^{BHM}-\Phi^{BM}$ strengthens the opinion
that HI play a significant, though weak, role during the deposition
process investigated here.

     In order to confirm that $B_3$ corresponding to deposition with HI
is significantly smaller than its ballistic counterpart, we have
developed a new simulation method to determine this coefficient, as
explained in the next subsection.

\subsubsection{Analysis at low coverage} \label{phitheta2}

 In the BM, at least three particles are required to form a trap for a
depositing particle.  The term $B_3^{BM}\theta^3$ appearing in the
series expansion of $\Phi^{BM}(\theta)$ precisely reflects the
rejection efficiency of three particle configurations leading to the
rejection of a new incoming one.  As already discussed by Thompson and
Glandt\cite{thompson1992} for the BM, an isolated triangle formed by
the centers of three adsorbed spheres is a trap if and only if (i) it
has no side longer than twice the particle diameter, (ii) all its
angles are acute, and (iii) the radius of its circumcircle is not
larger than one sphere diameter.  For an adsorbing square surface of
area $s$, the probability for an incoming particle to be rejected is
given by the ratio of the area of the triangle to $s$.  When the
triangle is not a trap, its exclusion area is evidently equal to zero.

     In the BM it is easy to build up a large number of representative
traps formed by 3 deposited particles and to evaluate their average
exclusion area $<A_{ex}>$.  Consider now a large adsorbing surface of
area $S$ covered with $N$ particles and virtually subdivided into a
large number $\nu$ of sub-surfaces of area $s$ ($\nu = S / s$).  The
probability $p_1$ that a given sub-system contains effectively three
particles is given to a good approximation by

\begin{equation}
p_1=\left(\begin{array}{c}3\\N\end{array}\right) \left(\frac{1}{\nu}\right)^3
\left(1-\frac{1}{\nu}\right)^{N-3}
				\label{p11}
\end{equation}

\noindent In this formula we assume that at low coverage the
sub-systems have no mutual influence, which is indeed correct.  The
probability $p$ for an incoming particle over the surface $S$ to be
trapped is then given by the product of the probability $p_3$ that it
deposits in a sub-system containing at least 3 particles and the
conditional probability $q$ that this particle which ends up over a
sub-system which is known to contain at least three adsorbed particles
is trapped by them. To lowest order in the coverage, $p_3$ is equal to
$p_1$  and the conditional probability $q$ is then given by
$<A_{ex}>/s$.  Thus one gets:

\begin{equation}
p\approx\left(\begin{array}{c}3\\N\end{array}\right)
\left(\frac{1}{\nu}\right)^3
\left(1-\frac{1}{\nu}\right)^{N-3}
\frac{<A_{ex}>}{s}			\label{p2}
\end{equation}

     The probability $p$ can be identified with $B_3^{BM}\theta^3$ (where
$\theta =\pi N/(4 S)$, provided that the diameter of the particles has
been taken as unity) in the density expansion of $\Phi$.  In the limit when
$N\rightarrow\infty, \nu\rightarrow \infty$, with $N/\nu\rightarrow
0$, it follows that:

\begin{equation}
 B_3=\frac{<A_{ex}> s^2}{3!  \left(\frac{\pi}{4}\right)^3}
\label{b3}
\end{equation}

     A simulation consisting in the deposition of $10^8$ independent
sets of three particles on a square surface of side length equal to
5,6,...,40 (in units of the diameter), hence of area ranging from
$S=25$ up to 1600, leads to $<A_{ex}> s^2=28.89\pm 0.16$.  Inserting
this value into eq.(\ref{b3}), one finds the value $B_3^{BM}=9.939\pm
0.055$, in good agreement with the theoretical value 9.94978 given by
Choi {\em et al.}\cite{choi1993}.  Hence, the method provides a
convenient means for estimating the first non-vanishing term of the
series expansion of the available surface function $\Phi$. It can also
be applied to the deposition process which takes HI into account.  In
this case, a triangle cannot be a trap if it does not constitute a trap
for the BM.  However, even though a triangle may act as a trap in the
presence of HI, its rejection efficiency is no longer proportional to
its geometrical area, but it can be significantly smaller.  It can
never be larger because, as already pointed out, HI introduce an
effective repulsion between the adhering particle and the preadsorbed
ones. Therefore, each side of the {\it ballistic} trapping triangle
becomes a concave curved line due to the repulsive hydrodynamic effect
of the particle located at the opposite vertex.  Also as a result of
this effective repulsion, the number of traps formed in the presence of
HI should be smaller than the ones formed in a pure ballistic
"experiment" for the same coverage.  The result of these various
effects can only be evaluated by simulation.  We have therefore
developed a special algorithm aimed at the construction of triangles in
the presence of HI.

	A large number of sets of three particles were deposited on
surfaces of area $s$ in the presence of HI.  In order to evaluate
$<A_{ex}>$, for each set of three particles the exclusion area $A_{ex}$
should be determined by depositing a fourth particle on the surface $s$
a large number $N_p$ of times.  $ A_{ex}/s$ is then given by the ratio
of the number of successful deposition trials of this fourth particle
to the total number of trials $N_p$.  $<A_{ex}>$ is then simply the
average of these exclusion areas $A_{ex}$ over the great number of
independent sets of three initial particles.  It can be noticed that
many of these sets lead to an exclusion area which is zero.  However,
this procedure to determine $<A_{ex}>$ is very time consuming from a
simulation point of view.  We have thus, in a first step, approximated
the exclusion area of a triangle by its geometrical area. We have
generated $10^5$ triangles taking HI into account, counting which
fraction of the generated triangles constitute a trap according to BM.
This leads obviously to an upper limit for $B_3^{BHM}$, estimated to be
approximately 7.7.  This rough approximation shows that in the presence of
HI, $B_3^{BHM}$ is at least 22\% lower than in the ballistic case. This
arises from the fact that, on average, the trapping triangles generated
by BHM are larger than the ones obtained in the BM due to the effective
repulsion of HI.

	In order to get a more precise estimate of $B_3^{BHM}$, we
should also take into account the change related to the decrease of the
excluding area for a given trap. However, the general simulation scheme
introduced in the previous paragraph is too time consuming, as already
mentioned. We have looked at a simpler procedure by studying first the
relative frequencies of the different trapping triangles.  We have
characterized a triangle by its largest angle and its area, and studied
the histogram of trapping triangles. In fig. 7(a-b), we have plotted
the histograms of relative frequencies for BM and BHM respectively,
constructed using the same procedure which has allowed us to give an
upper limit for $B_3$.  In both cases, one can see that the equilateral
touching triangles are the most probable objects, because the second
particle has rolled over the first one, and the third rolls over the
two preadsorbed ones.  Afterwards, there exists a relatively high curve
representing those triangles in which  the two latter adhering spheres
have rolled only over one preadsorbed particle. One can easily verify
that this curve behaves as $\sin (\alpha)/2$, $\alpha$ being the
largest angle of the triangle, since this is the area of such
triangles. Besides these singular contributions, there exists a
plateau, which corresponds to those triangles in which only one
adhering sphere has rolled over a preadsorbed particle.  Finally, the
remaining traps formed without rolling give a negligible contribution
to the histogram.  Although this general description applies for both
models, in the BHM the rolling mechanisms is not as effective, due to
the repulsion induced by HI. Nonetheless, we have assumed that by
studying only the most probable triangles, one can still improve our
first estimate of $B_3$. To this end, we have performed numerical
simulations in which we prescribe a rectangle of an area twice a given
trapping triangle, and we calculate its excluding area by the general
and rigorous method depicted in the previous paragraph. We fix the
triangle to have two sides of length 1, and the largest angle to be
larger than $60^{o}$, and for each given triangle, we analyze the
deposition of $10^5$ particles starting within the prescribed
rectangle, counting the fraction of such particles which are able to
reach the surface.

	As shown in table I, the trapping area remains almost constant
with the angle. If one compares the results of the BHM to BM, one can
clearly see that for these small triangles, the excluding area is
reduced to almost half its BM counterpart due to HI repulsions.  If we
take into account that the mean geometrical area of the triangles is
larger in the presence of HI, which reduces the fraction of traps, and
that in addition the excluding area is also reduced by the effective
repulsion, we get as a better estimate $B_3^{BHM}\approx
7.7\mbox{x}0.526=4.05$, where 0.526 is the ratio of the mean rejection
fraction in BHM (0.2634) to the mean rejection fraction obtained by BM
(0.5005), according to table I. We can then set bounds to this
coefficient, since it should obey,
$4.146\pm0.003<B_3^{BHM}<7.7\pm0.003$.  It is worth pointing out that
the lower bound is close to the one obtained by fitting $\Phi$ with a
power series as seen in subsection \ref{phitheta1}. There, we obtained
a value of 4.7.  Our new estimate has to be a lower estimate, since we
have disregarded the influence of the larger triangles, which also have
a larger exclusion area. Nonetheless, the lower estimate is quite close
to the value obtained by the fitting procedure, indicating that,
indeed, $B_3^{BM}/B_3^{BHM}\approx 2$.  This result shows that
HI strongly influence the local structure of the deposits, modifying
the triplet distribution.

	We have finally looked at the fraction of incoming particles
ending inside an equilateral triangle as a function of its side. Again,
for a given equilateral triangle, we let deposit $10^4$ particles and
calculate the fraction which ends within the triangle. As seen in fig.
8, even for triangles of side length 7 diameters, the fraction is
smaller than the value 0.5 predicted by BM. This again shows the
long-range character of HI, and its tendency to form
looser aggregates on the substrate with respect to BM predictions.

	 It would be interesting to determine the value of $B_3^{BHM}$
 more precisely but this requires long computer times.  Its study, as
well as the evolution of $B_3$ with $R^*$ is currently under way by
using the general and rigorous method presented here.  It will be the
purpose of a future article.

\begin{table}
\begin{center}
\begin{tabular}{ c  c  c }\hline
  angle  ({\it in degrees}) &  BM & BHM \\
 \hline \hline
	             60.0 & 0.4987 & 0.2587 \\
                     67.5 & 0.4995 & 0.2581 \\
	             75.0 & 0.5077 & 0.2694 \\
	             82.5 & 0.5003 & 0.2632 \\
	             89.9 & 0.4961 & 0.2676 \\
\hline 	                            \label{tabb1}
\end{tabular}

\caption{Fraction of incoming particles which are trapped on an area
which is twice the area of the corresponding rectangle triangle both for the
BM and BHM. The largest angle characterizing the trapping
triangle is given in the first column, and the shortest side is always
of one particle diameter.}

\end{center} \end{table}

\section{Conclusions}

	     This article presents a first study in which experimental
results concerning the deposition of large particles on a solid surface
under the influence of gravity have been compared to both the ballistic
deposition model and a simulation model which takes {\bf hydrodynamic
interactions} (HI) into account.  We have performed numerical simulations of
the deposition of spherical particles on a surface, incorporating the
effect of HI, in the case when Brownian motion can be neglected,
thus generalizing for a bidimensional surface the previous simulation
data for the deposition on a linear substrate\cite{ignacio1994}. The
major effect of HI is the induction of an effective repulsion between
the adsorbing sphere and the preadsorbed particles. We have then been
able to compare the simulation data with the pair distribution function
obtained experimentally for the deposition of melamine particles.  We
have shown that in the full range of surface coverages, the comparison
of the BHM with the experimental results leads to a better agreement
than the one obtained when comparing BM to the experimental data.  In
particular, the height of the first peak of $g(r)$ is improved,
especially at low coverages, when the fraction of rolling particles
plays an important role. The broadening of the curve after the first
peak, which the BM always underestimates, is also in better agreement.
These features appear as a result of the effective repulsion induced by
HI, although they lead to a quantitative small change in the curve
(except in the height of its first peak), because HI does not alter the
qualitative features of the adsorption process.  As a matter of fact,
HI will become more significant the closer we look at the structure of
the adsorbed layer.  In this sense, we have also studied the third
virial coefficient $B_3$ of the available surface function $\Phi$.  We
have shown that this coefficient drops to half its BM value. This
comes from the fact that, due to the effective repulsion, at a given
coverage, the fraction of triangles that form a trap is smaller in the
BHM, and that the excluding area of a trapping triangle is also smaller
than the geometrical area of the triangle. However, the decrease of
$B_3$ does not produce a significant change in $\Phi$. As a general
remark, we can conclude that HI have only a small effect on global
averaged quantities. In this respect, the BM constitutes a good
approximation, and this study thus validates the BM which has already
been widely studied from a theoretical point of view. On the other
hand, for a fine analysis of the local structure, one has to take HI
into account. It is not at all obvious, however, whether such
conclusions remain valid when the diffusion of the particles in the
bulk plays some role {\em i.e}.  for value of $R^*$ of the order or
smaller than 3.  In this case HI can again play a major role and this
should be investigated in the near future.

\section*{Notation}

\noindent BHM : Sequential adsorption model with hydrodynamic
interactions at large gravity and BM-like rules\\

\noindent BM  : Ballistic model \\

\noindent HI  : Hydrodynamic interactions\\

\noindent RSA : Random sequential adsorption model\\

\section*{Acknowledgements}

	The authors wish to thank  D. Bedeaux, G. Koper and E. Mann
for fruitful discussions.This work has been supported by the Commission
of the European Communities under grant SCI$^{*}$-CT91-0696 and by
CICYT (Spain), grant PB92-0895, as well as by an INSERM-CSIC project in
the framework of a France-Spain cooperation.

\appendix

\section*{Elements of the sphere-sphere friction tensor}

In this appendix we give the explicit expressions for the friction
coefficients
corresponding to a geometry where two equal-sized spheres are suspended in an
 infinite fluid.

    Jeffrey and Onishi\cite{jeffrey1984} computed the corresponding
friction tensors for any distance separating them. We will follow the
notation of eq.(\ref{1.3}), considered in the special case of the force
acting on one sphere in the presence of another one suspended in an
unbounded fluid, and concentrate on the tensor components of interest.
The two spheres have a radius unity and are separated a distance $r$.
We will call $\hat{e}$ the unit vector of the center-to-center
direction. Then, due to the symmetry, the matrices appearing in
eq.(\ref{1.3}) can be expressed as

\begin{eqnarray}
\left(\xi_{tt}\right)_{ij}&=&X_{11}^A e_ie_j+Y_{11}^A
(\delta_{ij}-e_ie_j)                    \label{1a.1}\\
\left(\xi_{tr}\right)_{ij}&=&Y_{11}^B\epsilon_{ijk}e_k    \label{1a.2}\\
\left(\xi_{rr}\right)_{ij}&=&X_{11}^Ce_ie_j+Y_{11}^C\left(\delta_{ij}-
e_ie_j\right)                                   \label{1a.3}
\end{eqnarray}

     \noindent where the different functions correspond to the
expressions appearing in eq.(\ref{1.6}), and the subindices correspond
to the different spatial direction. These new functions
are given by

\begin{eqnarray}
X_{11}^A&=&\frac{1}{4}\frac{1}{1-4
r^{-2}}-\frac{9}{40}\ln\left(1-\frac{4}{r^2}
\right)- \frac{3}{112}\left(1-\frac{4}{r^2}\right)
\ln\left(1-\frac{4}{r^2}\right)                 \nonumber\\
&+&\frac{3}{4}+\frac{17}{70}\frac{1}{r^2}+\frac{127}{560}
\frac{1}{r^4}-\frac{4057}{2240} \frac{1}{r^6}+...       \label{1a.4}\\
Y_{11}^A&=&-\frac{1}{6}\ln\left(1-\frac{4}{r^2}\right) +1-\frac{5}{48}
\frac{1}{r^2}+\frac{371}{768}\frac{1}{r^4}+\frac{16331}{r^6}+...\label{1a.5}\\
Y_{11}^B&=&-\frac{1}{4}\left[1+\frac{1}{2} \left( 1-\frac{4}{r^2} \right)
\right] \ln\left(\frac{r+2}{r-2}\right) +\frac{3}{2r}
-\frac{9}{8 r^3}+\frac{1}{1920 r^5}+... \label{1a.6}\\
Y_{11}^C&=&-\frac{1}{5}\left[1+\frac{47}{50} \left(1-\frac{4}{r^2}\right)
\right] \ln\left(1-\frac{4}{r^2}\right)+1-\frac{194}{125}\frac{1}{r^2}
                                                \nonumber\\
&+&\frac{327}{500}\frac{1}{r^4}-\frac{51853}{24000}\frac{1}{r^6}+...
                                        \label{1a.7}
\end{eqnarray}

\noindent In these expressions the Stokes value for the friction
     coefficient is taken as unity.


\newpage
{\bf Figure captions}

{\bf Fig. 1} Motion of an adsorbing particle close to a preadsorbed one, when
they are at a distance $r$, with a clearance $\Delta r$. The time increment in
the numerical algorithm is chosen so that $\Delta r/r\leq 0.1$. As a result,
due
to the anisotropic behavior of the mobility, the motion of the sphere is almost
parallel to the surface of the adsorbed sphere, {\em i.e.} $\Delta r_p >>\Delta
r$. Therefore, the trajectory of the adsorbing particle will approximately
follow the dotted line, which corresponds to the trajectory predicted by  {\it
BM}.\\

{\bf Fig. 2} {\it Interaction cylinder}: (---) Constant-radius interaction
cylinder, and (- - -) variable-radius interaction cylinder for an adsorbing
sphere.  The particle is left at an initial height $h_0$ and in the second
case,
the radius of the cylinder diminishes as the height until a height of 5
diameters is reached. At that point the cylinder is kept at a constant radius.
In the first case, the adsorbing particle will interact with all the dark
adsorbed spheres; in the second, at the end only the black one will influence
its motion.\\

{\bf Fig. 3} One surface covered by spheres using a varying-radius (o), and a
constant-radius {\it interaction cylinder} ($\bullet$) (see fig. 2), at a
coverage of
a) $\theta=0.25$, and b) $\theta=0.5$ (the jamming corresponds approximately to
0.6).\\

{\bf Fig. 4} Pair distribution functions:  Comparison of the experimental data
($\bullet$) with the results of a) BM and b) BHM respectively at
$\theta=0.1489$. c) and d ) Equivalently at $\theta=0.3495$. e) and f)
Correspondingly at $\theta=0.4997$.\\

{\bf Fig. 5} a) Available surface function, $\Phi (\theta)$, for BM (---) and
BHM (o) as a function of the coverage $\theta$ obtained after covering
surfaces of area 23.34x27.34 diameters. The results are obtained after covering
400 surfaces in the BM, and 500 surfaces in BHM until $\theta=0.30$
and 300 from $\theta=0.30$ up to $\theta=0.50$. b) Same as Fig. 5a, except that
the sample size is 5000 surfaces for BHM, and the surfaces are covered until
$\theta=0.1$.\\

{\bf Fig. 6} Difference between the surface available function of BM
and BHM. Note that the average is displaced towards the region of positive
numbers.\\

{\bf Fig. 7} Histograms of the relative frequency of the trapping triangles
(see
definition in the text) as a function of the area and maximum angle of such
triangles for a) BM, and b) BHM.\\

{\bf Fig. 8} Fraction of particles adsorbed inside an equilateral triangle
of side length $l$ if initially they are randomly chosen on a rectangle
of area twice the corresponding triangle area. (---) BHM, (- - -) BM.

\end{document}